\documentclass[letterpaper,12pt]{article}
\pdfoutput=1

\usepackage{jheppub}
\usepackage{amsmath}
\usepackage{graphicx}
\usepackage{subfig}
\usepackage{cancel}
\usepackage[dvipsnames]{xcolor}
\usepackage{color}
\usepackage{mathrsfs}
\usepackage{amsfonts}
\usepackage{dsfont}

\allowdisplaybreaks

\def\({\left(}
\def\){\right)}
\def\[{\left[}
\def\]{\right]}

\newcommand{\be}{\begin{equation}}
\newcommand{\ee}{\end{equation}}

\newcommand{\ban}[1]{\begin{align}#1\end{align}}

\newcommand{\ket}[1]{\left| #1\right>}
\newcommand{\bra}[1]{\left< #1\right|}

\title{Living on the Edge: A Toy Model for Holographic Reconstruction of Algebras with Centers}

\author{William Donnelly, Ben Michel, Donald Marolf, and Jason Wien}

\affiliation{Department of Physics, University of California, Santa Barbara, CA 93106, USA}

\emailAdd{donnelly@physics.ucsb.edu}
\emailAdd{michel@physics.ucsb.edu}
\emailAdd{marolf@physics.ucsb.edu}
\emailAdd{jswien@physics.ucsb.edu}

\abstract{
We generalize the Pastawski-Yoshida-Harlow-Preskill (HaPPY) holographic quantum error-correcting code to provide a toy model for bulk gauge fields or linearized gravitons.  The key new elements are the introduction of degrees of freedom on the links (edges) of the associated tensor network and their connection to further copies of the HaPPY code by an appropriate isometry.  The result is a model in which boundary regions allow the reconstruction of bulk algebras with central elements living on the interior edges of the (greedy) entanglement wedge, and where these central elements can also be reconstructed from complementary boundary regions.   In addition, the entropy of boundary regions receives both Ryu-Takayanagi-like contributions and further corrections that model the $\frac{\delta \text{Area}}{4G_N}$ term of Faulkner, Lewkowycz, and Maldacena.   Comparison with Yang-Mills theory then suggests that this $\frac{\delta \text{Area}}{4G_N}$ term can be reinterpreted as a part of the bulk entropy of gravitons under an appropriate extension of the physical bulk Hilbert space.
}

\makeatletter
\def\@fpheader{\relax}
\makeatother

\begin{document}
\maketitle

\section{Introduction}

Recent works \cite{almheiri, happy, haydenRTN, harlow} have introduced models of gauge/gravity duality based on quantum error correcting codes and thus provided a new paradigm for studying holographic systems.  The models implement their codes via tensor networks that map bulk logical operators to operators on a code subspace of a larger boundary Hilbert space. Such representations were termed ``holographic codes" in \cite{happy} and have been shown to exhibit key properties of the AdS/CFT correspondence such as bulk reconstruction and the Ryu-Takayanagi (RT) relation between entanglement in the boundary theory and the area of bulk minimal surfaces \cite{rt1,rt2}.

Indeed, as noted in \cite{happy}, such holographic codes also reproduce an important part of the $1/N^2$ corrections to RT found by Faulkner, Lewkowycz, and Maldacena  (FLM) \cite{Faulkner:2013ana}.  Recall \cite{Faulkner:2013ana} that with such corrections the entropy $S_A$ of a boundary region $A$ takes the interesting form
\be
\label{eq:flm}
S_A = \frac{\text{Area}}{4G_N}+S_\text{bulk}(\rho_{W(A)}) + \frac{\delta \text{Area}}{4G_N} + \dots
\ee
The first term on the right is the leading-order Ryu-Takayanagi piece, which is local on the entangling surface and independent of the state. The second accounts for bulk entropy in the entanglement wedge $W(A)$ defined by the RT minimal surface, and is thus generally both non-local and non-linearly dependent on the bulk state. The third is an additional effect from quantum corrections to the RT area, which is distinguished by being both local on the RT surface and linear in the bulk state; i.e. it is an expectation value. The $\dots$ denote higher order terms in the $1/N$ expansion.  In the codes from \cite{happy}, the analogous result contains the first two terms on the right-hand side. 

It was suggested in \cite{harlow} that the remaining $\frac{\delta \text{Area}}{4G_N}$ term would also arise naturally from a quantum error correction model containing operators ${\cal O}$, associated with the boundary between the entanglement wedge of $A$ and that of its complement $\bar A$, that are reconstructible from both $A$ and $\bar A$. Such ${\cal O}$ must lie in the center of either reconstructed algebra.  The terms $\frac{\text{Area}}{4G_N}+\frac{\delta \text{Area}}{4G_N}$ then naturally correspond to aspects of the code that are, in some sense, dependent on the values of operators in this center.

Much of the above structure is familiar from analyses \cite{Buividovich,will_lattice,Casini} of entropy in lattice gauge theories.
In that context, the (electric \cite{Casini}) algebra of operators acting in a bulk subregion contains the electric fields $E_{\ell}|_{\partial A}$ along the links $\ell$  at the boundary of the subregion \cite{Buividovich, will_lattice, Casini,will14}.  And since Gauss's law equates the $E_{\ell}|_{\partial A}$  with operators spatially separated from $A$, the boundary electric fields commute with the entire subalgebra on $A$.   In particular,  the canonical conjugates of the $E_{\ell}$ are closed Wilson loops that pass through $\ell$, which are not elements of the subalgebras on either $A$ or $\bar A$ when $\ell\in\partial A$.

It is thus natural to consider modifications of the HaPPY code inspired by lattice gauge theory and having additional degrees of freedom that live on the links of the bulk lattice.  This is done in section \ref{section:construction} building on the HaPPY pentagon code \cite{happy}.  As desired, a key feature of our model is the existence of bulk operators that are reconstructible on both a boundary region and its complement.  Such properties are derived in section \ref{section:reconstruction} and follow directly from results of \cite{happy}.  We then demonstrate in section \ref{section:algent} that such central elements do indeed endow our model with an FLM-like relation containing analogues of all three terms shown explicitly on the right-hand side of \eqref{eq:flm}.  Section \ref{section:discussion} concludes with some final discussion.  In particular, comparison with lattice gauge theory constructions suggests that the FLM $\frac{\delta \text{Area}}{4G_N}$ term might be usefully reinterpreted as part of the bulk entropy of metric fluctuations in an appropriate extension of the physical bulk Hilbert space.

\section{Edge Mode Construction}
\label{section:construction}

The fact that gauge theories are described canonically by a connection and a conjugate electric flux makes it natural to describe these degrees of freedom as living on the links of a discrete graph-like model, as is common in lattice gauge theory.  This allows holonomies to be described as paths through the lattice and the Gauss law constraint to be imposed by requiring the electric fields on links attached to any vertex $v$ to sum to the charge at $v$.

Since we wish to extend holographic codes in a manner reminiscent of bulk gauge theories, we will introduce degrees of freedom below on the links of the tensor network corresponding to the pentagon code of \cite{happy}.  We will first review the relevant features of this code and then describe the desired augmentations.

\begin{figure}[h!]
\centering
\subfloat[]{\includegraphics[width=0.35\textwidth]{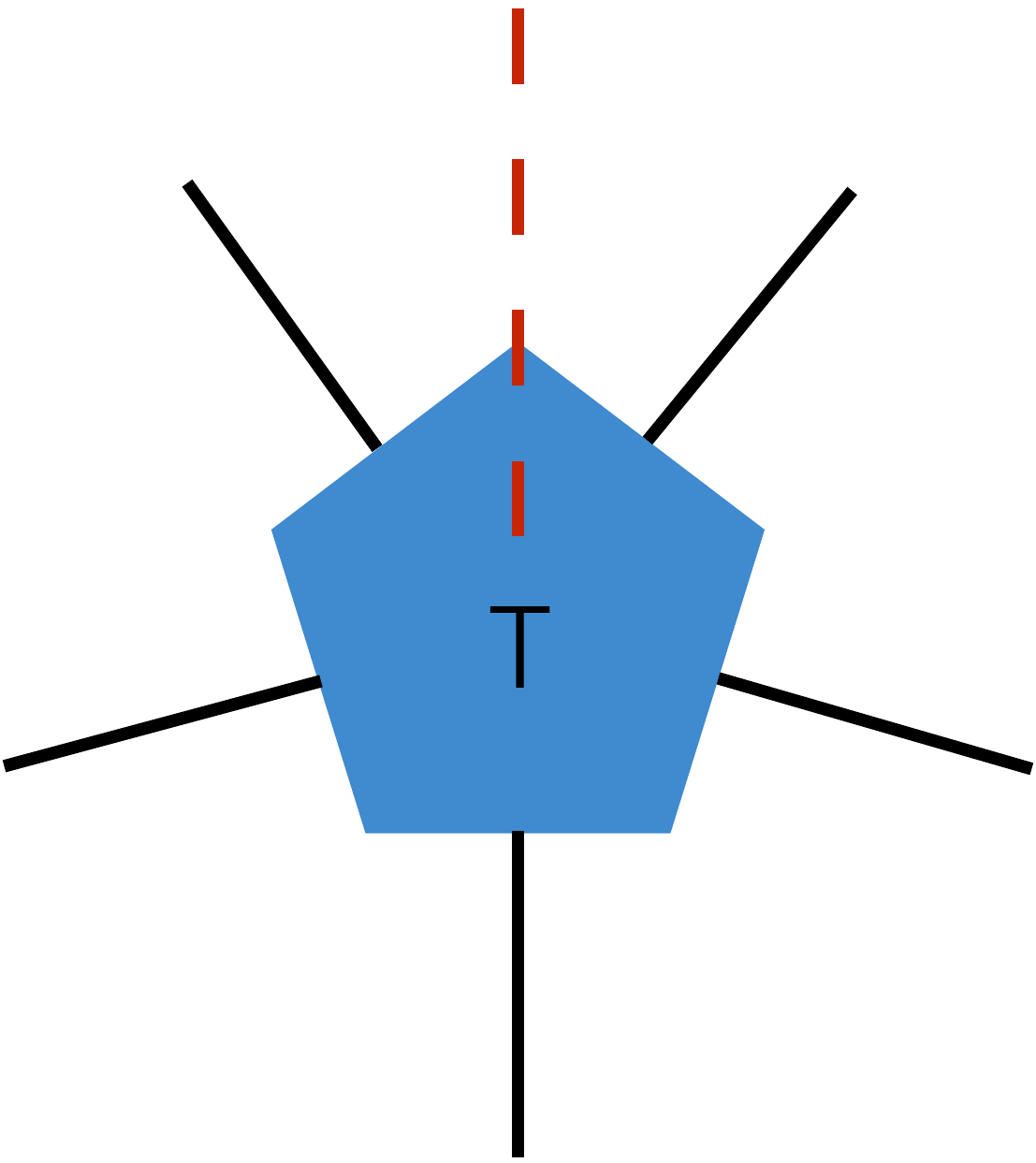}}\qquad
\subfloat[]{\includegraphics[width=0.35\textwidth]{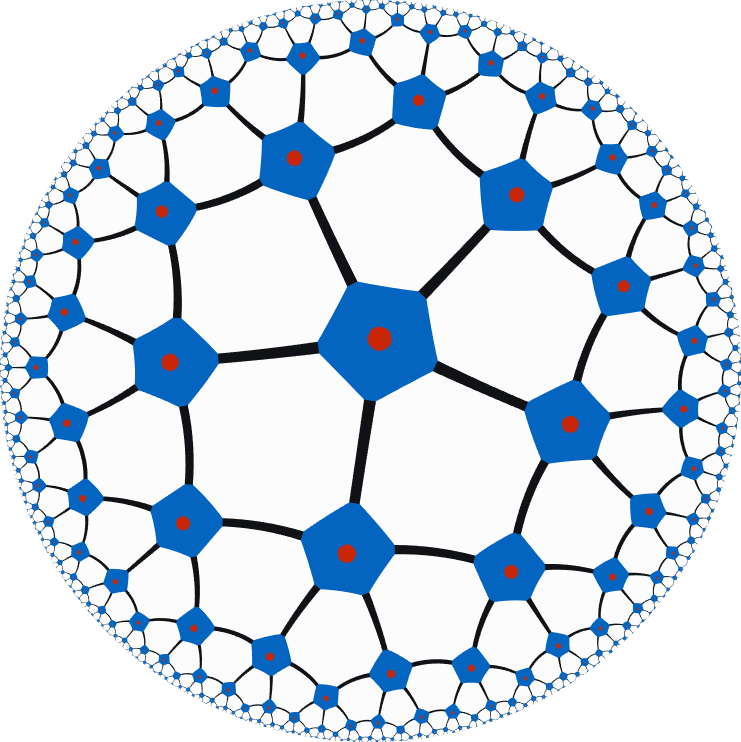}}
\caption{(a) The fundamental tensor $T$ of the pentagon code showing the bulk leg (dashed line, red in color version) and the network legs (solid lines). (b) These units are contracted along their networks legs to form a pentagonal tiling of the hyperbolic plane.
}
\label{pentagoncode}
\end{figure}

The pentagon code is a tiling of a hyperbolic disk where the fundamental unit is a six index tensor $T$ drawn in figure \ref{pentagoncode}.  The disk has finite size as the code is to be thought of as a model of a holographic CFT with a cutoff. Except at the boundary of the disk, five of the legs of the tensor are connected to adjacent tensors as depicted in figure~\ref{pentagoncode}. Even at the boundary, we refer to these five as network legs.  Each such tensor has one uncontracted index representing a local bulk degree of freedom. If $T$ is chosen to be a perfect tensor, meaning that it describes an isometry from any subset of at most 3 legs to the rest, an operator $\mathcal{O}$ acting on any bulk input can be ``pushed'' along three of the output legs to three adjacent tensors: the action of $\mathcal{O}$ on $T$ can be replaced by the action of $\mathcal{O}'=T^{\dagger} O T$ on one of the adjacent tensors.

This procedure allows us to push local bulk operators to the boundary, as the negative curvature of the hyperbolic plane ensures that each tensor has at least three legs pointing toward the boundary in a suitable sense. Since $T$ is a perfect tensor, one can also show \cite{happy} that the entropy is given by an FLM-like formula having analogues of the first two terms on the right-hand side of \eqref{eq:flm}.

\begin{figure}[h!]
\centering
\includegraphics[width=0.3\textwidth]{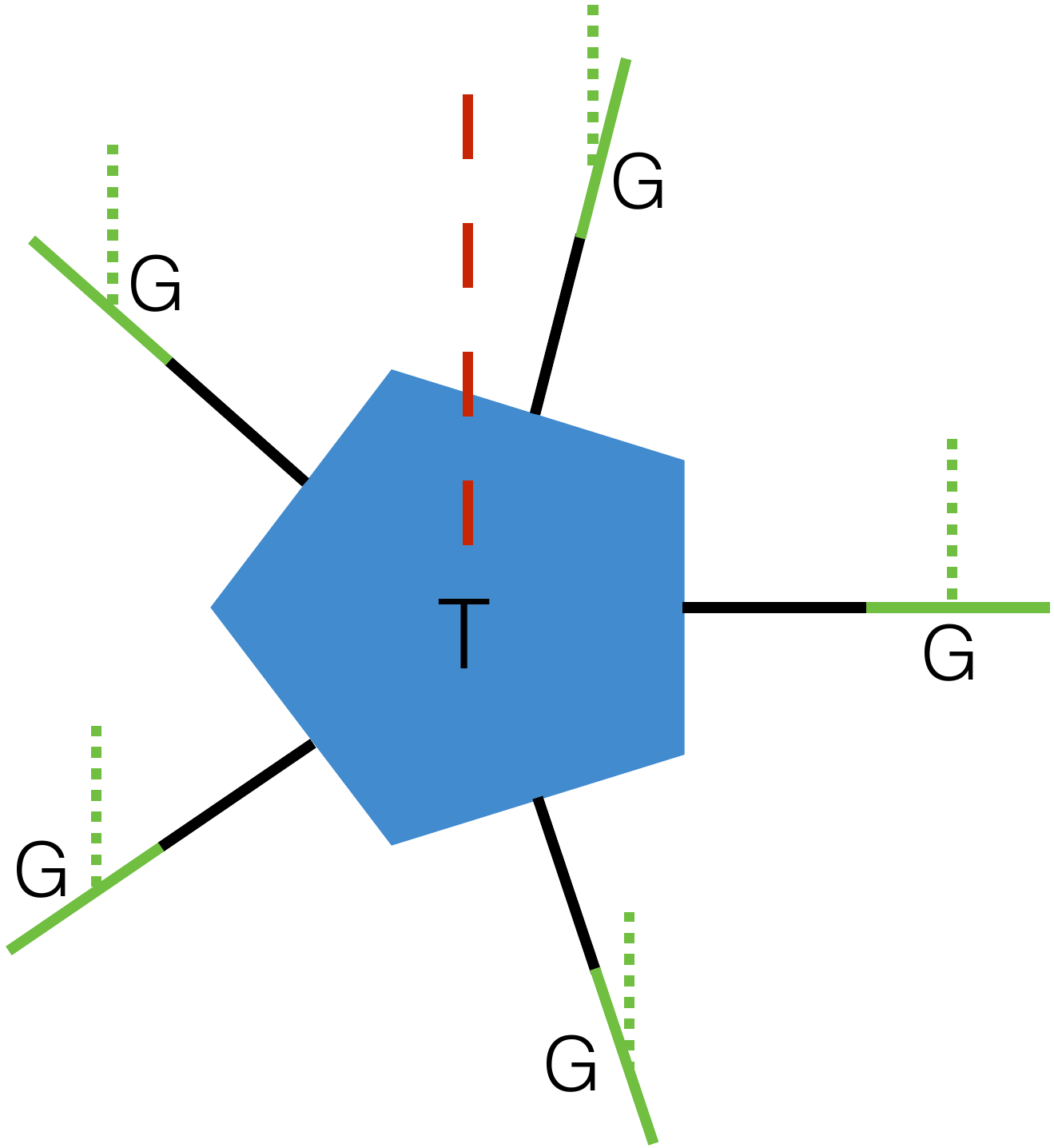}
\caption{An unsuccessful first attempt to add edge degrees of freedom.
A copy of the tensor $G$ has been attached to each of the 5 network legs of the tensor $T$ from figure \ref{pentagoncode}.  The bulk input leg of $G$ is drawn in small dashes. This attempt does not succeed, as the tensor annihilates bulk states lacking particular correlations among the 6 bulk inputs.}
\label{nogo}
\end{figure}

We wish to introduce additional degrees of freedom modeling bulk gauge fields in a way that largely preserves these properties.  As a first guess, one might add to each of the non-bulk legs of the fundamental unit a three index tensor $G_{ijk}$, whose role in the network is to link two adjacent tensors to a common input modelling the electric flux of some bulk gauge field.  One might then choose the tensor structure
\begin{equation}
\label{simpleG}
G=\delta_{ij}\delta_{jk}\delta_{ik}
\end{equation}
to impose flux conservation along each each link in the network. This new fundamental unit is drawn in figure \ref{nogo}.
However, the values on all the network legs are then determined by the inputs to the associated $G$s, so there is no room for further input from the bulk leg of $T$.  Indeed, the network just described will annihilate all bulk states orthogonal to a space in which the $T$ inputs are determined by the $G$ inputs (and where the $G$ inputs also satisfy a further set of constraints).

This unfortunate issue can be resolved by considering a model in which the above 6 bulk inputs are manifestly independent.  We do so by extending the fundamental unit $T$ to the 6-fold tensor product $\otimes_{m=1}^6 T =  T\otimes T\otimes T\otimes T\otimes T\otimes T$ and again connecting these units as in the pentagonal tiling of the hyperbolic disk shown in figure \ref{pentagoncode} (b).  Each factor in the resulting tensor product will be called a ``copy'' of the network: the first copy will be treated as an independent HaPPY network, while the additional copies will be contracted with our 3-legged tensor (or indeed any isometry) $G$  as described below.

\begin{figure}[h!]
\centering
{\includegraphics[width=0.55\textwidth]{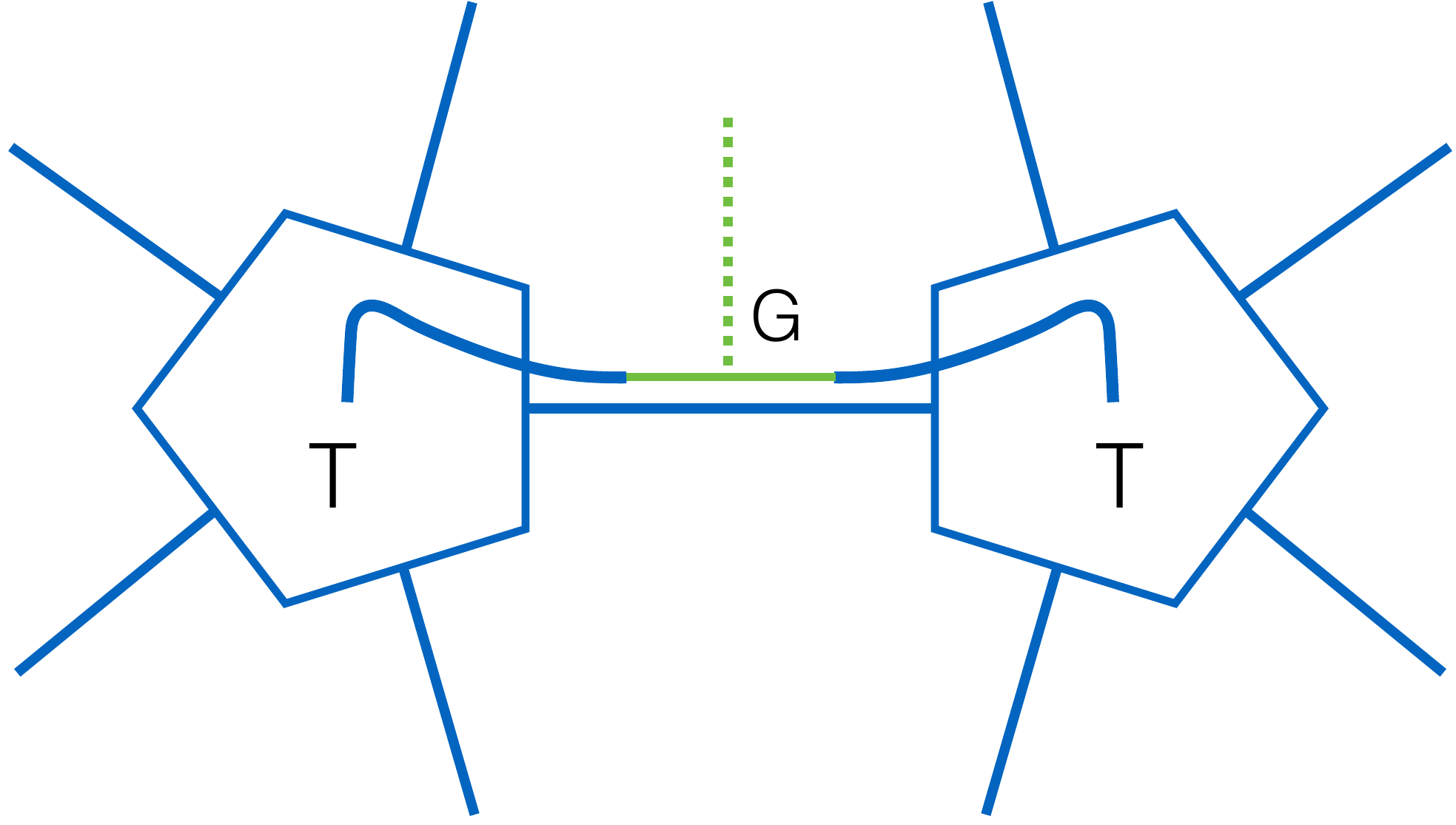}}
\caption{Our code is built from 6 copies of the code from \cite{happy} by contracting the tensor $G$ with a pair of neighboring bulk inputs. The relevant two $T$-tenors are shown here, where we have chosen them both to be part of the same copy of the pentagon code.}
\label{fund}
\end{figure}

\begin{figure}[h!]
\centering
\subfloat[]{\includegraphics[width=0.35\textwidth]{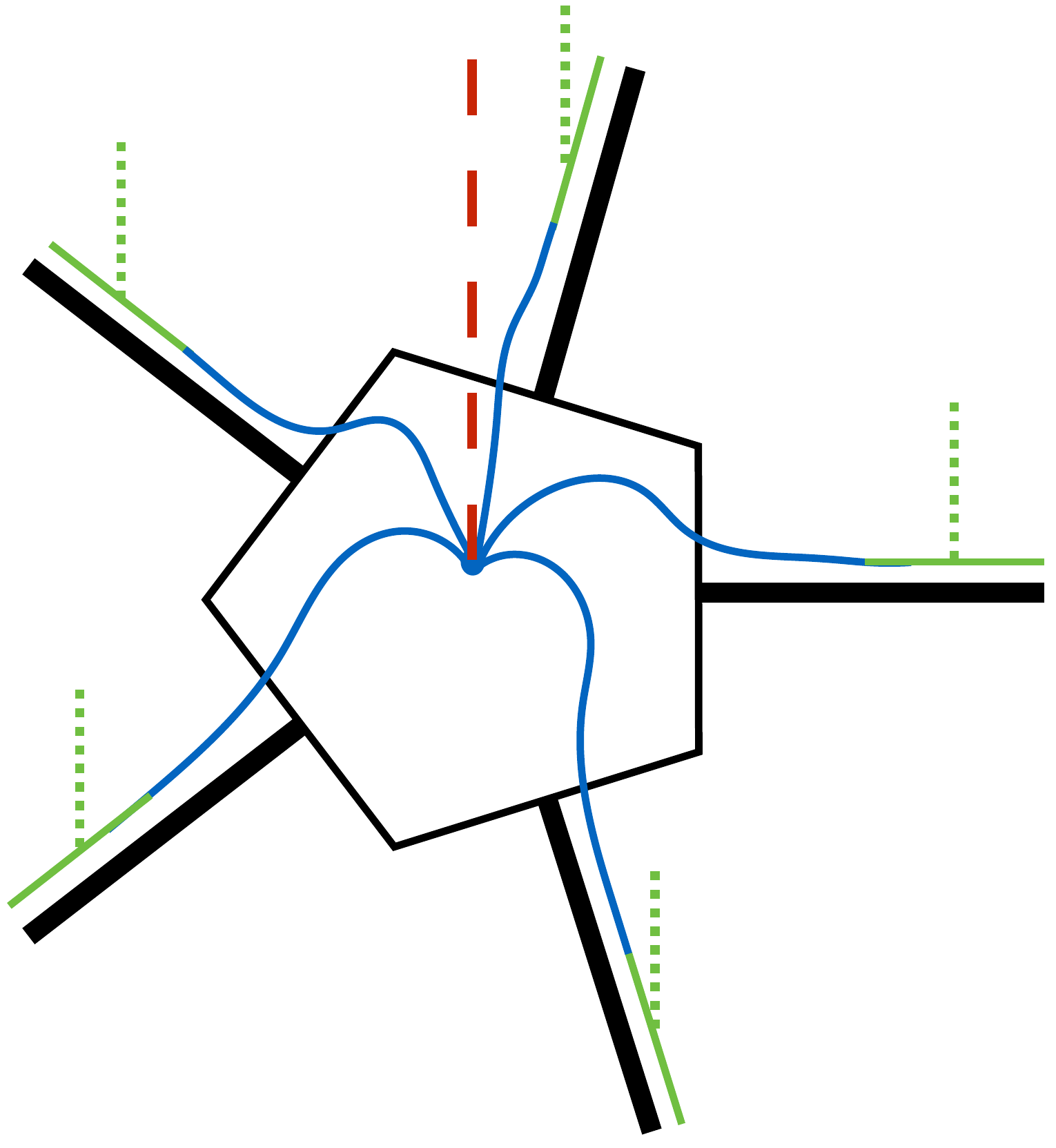}}\qquad
\subfloat[]{\includegraphics[width=0.35\textwidth]{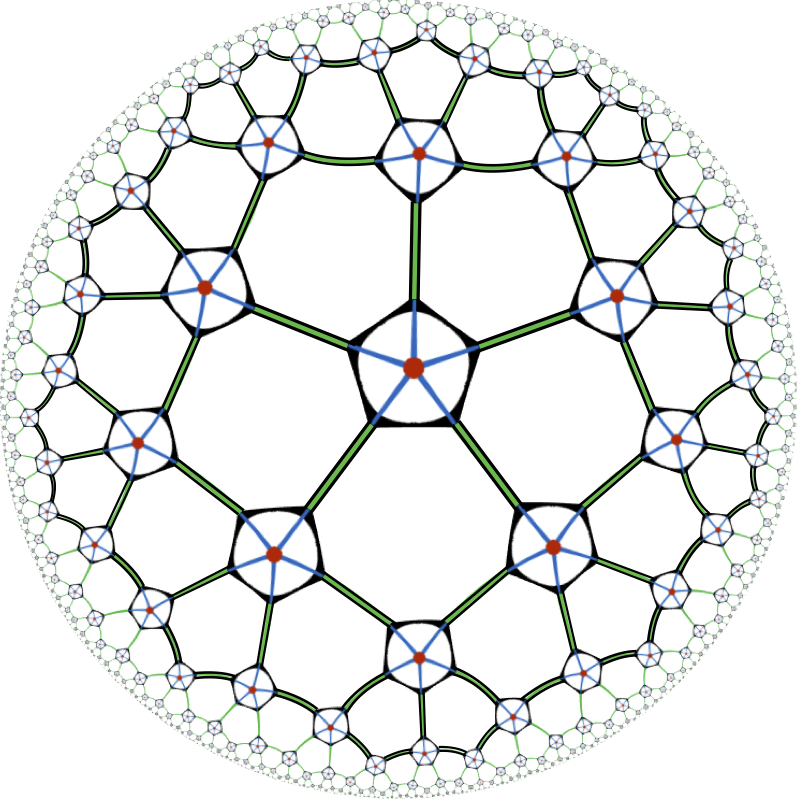}}
\caption{ (a) The structure near each vertex of our edge-mode code.  The thick black legs carry 5 indices.  The central input (long dashes, red in color version) corresponds to a bulk matter field as in \cite{happy} while the inputs on each edge (short dashes, green in color version) are to be interpreted as degrees of freedom of a bulk gauge field.  (b) A sketch of the full edge-mode code.
}

\label{edgecode}
\end{figure}

Thus far our network has $6$ bulk input legs at each vertex.  We will turn $5$ of these into inputs associated with edges instead. Consider some particular edge in the interior of the disk and choose one input leg from each of the two vertices it connects (to simplify the figures, both input legs are chosen from the same copy).  Our edge-mode code is constructed by contracting these legs with two legs of the tensor $G$; see figure \ref{fund}.  We will treat these two legs of $G$ as output legs; the remaining input is naturally associated with the edge under consideration.  Doing so for each edge uses $5$ of the bulk legs at each vertex, leaving the $6$th free to serve as a normal bulk input at each vertex just as in the original code from \cite{happy}.  To be concrete, we take this $6$th bulk input to live in the first copy of the network.  Figure \ref{edgecode} shows a pictorial representation of the full edge-mode code including all six copies the pentagon code.
Note that we have added one $G$ for every two bulk legs, and thus also for every two $T$s.

The resulting code defines an isometry from the bulk degrees of freedom to the boundary, and therefore has many of the same features as the code described in \cite{happy}. This is because one may view this edge-mode code as six copies of the HaPPY pentagon code together with a set of $G$ tensors interposed between these codes and the bulk state.  Since the $G$ tensors are isometries, composing them in this way with the original HaPPY network yields another isometry. As described below, this observation allows us to import all of the main technology from \cite{happy} including operator pushing, the greedy entanglement wedge construction, and the Ryu-Takayanagi formula for entanglement entropy. However, the additional tensors $G$ introduce certain subtleties which we will discuss in depth.


\section{Operators in the Center and Bulk Reconstruction}
\label{section:reconstruction}

We now consider properties of our code associated with subregion duality, showing that our model leads to the bulk reconstruction of algebras with centers.  This reproduces the structure suggested in \cite{harlow}.   Here we view the tensor network of our edge-mode code as an isometry from a bulk Hilbert space $\mathcal H_\text{bulk}$ defined by the set of all bulk inputs (both edge and vertex) to a boundary Hilbert space $\mathcal H_\text{bndy}$ defined by the set of network links that reach the boundary of the hyperbolic disk.  For simplicity of notation, we follow standard practice and use the above isometry to identify $\mathcal H_\text{bulk}$ with its image $\mathcal {\mathcal H}_\text{code}$ in $\mathcal H_\text{bndy}$.    Bulk operators are then maps from ${\mathcal H}_\text{bulk} ={\mathcal H}_\text{code}$ to itself.

As in \cite{almheiri,happy,harlow}, we shall say that a bulk operator $\mathcal O$ lies in the algebra ${\mathcal M}_A$ that
can be reconstructed from a region $A$ of the boundary if (and only if) there exists an operator $\mathcal O^{(A)}$ with support in $A$ such that
 \ban{
\mathcal O^{(A)} \ket \psi = \mathcal O \ket \psi \hspace{1cm} \forall \, \ket \psi \in \mathcal {\mathcal H}_\text{code} \, .
 }
Note that, with this definition, bulk operators ${\mathcal O}_1 \in {\mathcal M}_{A_1}$ and $\mathcal O_2 \in {\mathcal M}_{A_2}$ for non-intersecting regions $A_1$ and $A_2$ must commute.  In detail, on ${\mathcal H}_\text{code}$ we have
 \ban{
 \left[\mathcal O_1, \mathcal O_2 \right] \ket \psi & = \mathcal O_1^{(A_1)} \mathcal O_2 \ket \psi - \mathcal O_2^{(A_2)} \mathcal O_1 \ket \psi \notag \\
 & = \mathcal O_1^{(A_1)} \mathcal O_2^{(A_2)} \ket \psi - \mathcal O_2^{(A_2)} \mathcal O_1^{(A_1)} \ket \psi \notag \\
 &= \left[\mathcal O_1^{(A_1)}, \mathcal O_2^{(A_2)} \right] \ket \psi =0.
 }
 In the first step we have used the fact that bulk operators preserve ${\mathcal H}_\text{code}$, while the final step uses the fact that all operators in $A_1$ commute with those in $A_2$.  As a result, any bulk $\mathcal O$ lying in both ${\mathcal M}_{A_1}$ and ${\mathcal M}_{A_2}$ must be a central element of both algebras.

This is precisely the structure suggested by \cite{harlow} as the natural quantum-error-correction model of FLM corrections to the Ryu-Takayanagi relation. It is useful to contrast this situation with that of the HaPPY code, where reconstruction on $A$ succeeds for any operator in the greedy entanglement wedge $w^*(A)$ (or greedy wedge for short) defined by the greedy algorithm of \cite{happy}.\footnote{The greedy wedge associated with a region $A$ on the boundary is constructed by first taking all tensors with at least three legs contained in $A$. Next all tensors with at least three legs contracted with this set of tensors are included, and this procedure continues until there are no more tensors to add.} The boundary of this greedy wedge consists of two parts, one lying on the boundary of our hyperbolic disk and the other in the interior of the disk.  We refer to the latter as the greedy entangling surface $\gamma_A^*$. In our edge-mode code, we define a corresponding greedy wedge and $\gamma_A^*$ using only the tensors $T$ associated with our $6$ copies of the pentagon code.  When we then add the additional $G$ tensors, some bulk operators ($G$-inputs) act on links that straddle the resulting $\gamma_A^*$.  Only certain bulk operators acting on such links can be reconstructed on $A$, and those operators will typically lie in the center of the algebra of operators on $A$.

The essential point can be illustrated by considering only a pair of $T$s ($T_{L/R}$) that are linked by a single $G$ as in figure \ref{fund}.  We take the one bulk edge input to be a single qubit that feeds into $G$, and we take $G$ to map
\begin{equation}
\label{eq:copy}
\ket{0}\rightarrow \ket{00} \ \ \text{and} \ket{1}\rightarrow \ket{11}
\end{equation}
as in \eqref{simpleG}. For this reason we refer to $G$ as the copying tensor below.  The perfect tensors $T_{L/R}$ each have 4 uncontracted legs which we treat as  proxies for the left and right halves of the boundary.

\begin{figure}[h!]
\centering
\includegraphics[width=0.75\textwidth]{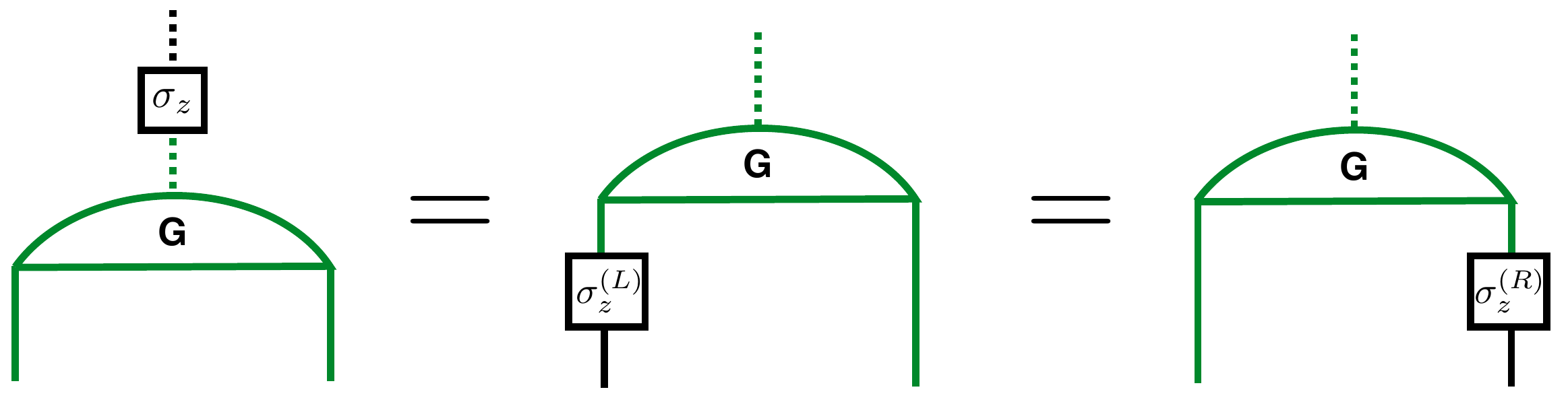}
\caption{Pushing $\sigma_z$ through $G$
}
\label{zpushing}
\end{figure}

One bulk operator of interest is the Pauli $\sigma_z$ defined by $\sigma_z |0\rangle= - |0\rangle, \ \sigma_z |1\rangle= |1\rangle$ acting on the bulk edge input. The structure of $G$ allows one to push $\sigma_z$ through $G$ onto either output leg of $G$: writing $G = \ket{00}\bra{0}+\ket{11}\bra{1}$, it follows that
\begin{equation}
\label{eq:toysigmaz}
G \sigma_z = -\ket{00}\bra{0} + \ket{11}\bra{1} = \sigma_z^{(L)} G= \sigma_z^{(R)} G.
\end{equation}
Here $\sigma^{(L/R)}_z$ denotes a corresponding Pauli matrix acting on the output of $G$ that feeds into $T_{L/R}$ as depicted in figure~\ref{zpushing}.
It follows that we can reconstruct $\sigma_z$ in the left boundary by pushing $\sigma_{z}^{(L)}$ through $T_L$. But the same procedure allows us to reconstruct $\sigma_z$ as an operator acting only on $R$ by pushing $\sigma_{z}^{(R)}$ through $T_R$. So as above $\sigma_z$ must lie in the center of ${\mathcal M}_L$, and also of ${\mathcal M}_R$.

As a result, the other Pauli operators $\sigma_x$, $\sigma_y$ at our bulk edge input cannot be reconstructed from either $L$ or $R$ alone.  But these operators can still be reconstructed if we are granted simultaneous access to both sets of boundary sites. Indeed, $\sigma_x$ satisfies
\begin{equation}
G\sigma_x = |00\rangle \langle 1| + |11\rangle \langle 0| = \sigma_x^{(L)} \sigma_x^{(R)} G.
\end{equation}
We may then push each of $\sigma_x^{(L/R)}$ through its respective $T$ to the boundary, and so any boundary region whose greedy entanglement wedge includes both $T_L$ and $T_R$ will be able to reconstruct the $\sigma_x$ that acts between them. $\sigma_y$ will also be reconstructible on the region, and $\sigma_z$ will not be central.

Returning to the full edge-mode code, we may consider the greedy wedge for any region $A$ on the boundary.  Bulk operators in this wedge may be generated by taking sums and products of the following three types of `local' operators:  (i) operators that act on inputs at a single vertex, (ii) operators acting on a single link that lies in the interior of the greedy wedge, and (iii) operators acting on a single link that straddles the corresponding greedy entangling surface.

Operators of type (i) are precisely the bulk operators defined in \cite{happy} and act on the first copy of the code constructed in section \ref{section:construction}, so as in \cite{happy} such operators lie in ${\mathcal M}_A$.  Operators of type (ii) were shown above to be equivalent to a pair of operators acting on tensors $T$ on either side of the link, thus they also lie in the interior of the greedy wedge.  In particular, each member of the pair acts on the leg of $T$ that was interpreted in \cite{happy} as a bulk input of a pentagon code.  So again such operators lie in ${\mathcal M}_A$.

For operators of type (iii), there are two cases.  When the operator acts like $\sigma_z$ above it can be replaced by a single operator acting on the adjacent tensor $T$ lying inside the greedy wedge. It acts on a leg that was interpreted in \cite{happy} as a bulk input of a pentagon code and so can be reconstructed in $A$.    However, other operators on this edge input cannot generally be reconstructed in $A$.  Indeed, when the relevant edge $e$ also straddles the greedy entangling surface of the complementary ($\bar A$) boundary region\footnote{For the entanglement wedges defined by minimal surfaces in gauge/gravity duality, this would always be true in a pure state as the entanglement wedge of $A$ is the complement of that for $\bar A$.  In contrast, in the model of \cite{happy} there can be a region that lies in neither the greedy wedge for $A$ nor that for $\bar A$. The existence of such a region is to be regarded as an artifact of the model associated with discretization of the bulk spacetime; see section \ref{section:entropy} for further discussion.} it follows as above that $\sigma_z$ lies in the center of both ${\mathcal M}_A$ and ${\mathcal M}_{\bar A}$.

We close this section by noting that our model admits a broad class of generalizations preserving the above properties.  The point is that the above arguments depended only on $G$ copying the input into both outputs. In particular, this makes $G$ and isometry from any one leg to the remaining two, allowing us to push operators of type (ii) as above to a two site operator completely contained in the greedy wedge. This remains the case if we break the symmetry between the first (input) leg of $G$ and the output legs (second and third) by  replacing \eqref{eq:copy} with any map of the form
\begin{equation}
\label{eq:gencopy}
G: H_\text{in} \rightarrow (H_\text{in} \otimes H_\text{aux})_L \otimes (H_\text{aux} \otimes H_\text{in})_R, \ \ \ G|\alpha \rangle \mapsto |\alpha\rangle | \psi(\alpha)\rangle |\alpha \rangle.
\end{equation}
Here $\{|\alpha \rangle \}$ is a basis for the input Hilbert space $H_\text{in}$ (which we call the copying basis), $| \psi(\alpha)\rangle$ is a  state on an auxiliary product Hilbert space $H_\text{aux} \otimes H_\text{aux}$, and the tensor factors marked $L,R$ in \eqref{eq:gencopy} correspond respectively to the two output legs of $G$. As a concrete example, one may consider

\be
\label{eq:gravitoncode}
G: \ket{0}\rightarrow \ket{0000},\quad\quad \ket{1}\rightarrow \ket{1}\otimes \frac{1}{\sqrt{2}}\left(\ket{00}+\ket{11}\right)\otimes \ket{1},
\ee
which has 

\be
\ket{\psi(0)} = \ket{00},\quad\quad \ket{\psi(1)}= \frac{1}{\sqrt{2}}\left(\ket{00}+\ket{11}\right).
\ee

The generalization \eqref{eq:gencopy} allows us to make direct contact with the extended-lattice discussion \cite{will14} of entropy in lattice gauge theories.  Ref. \cite{will14} considered regions of a lattice with the boundary $\gamma$ of the region taken to intersect only links (i.e., no vertices lie on $\gamma$).  Each link was then divided into two parts, with separate Hilbert spaces defined on the parts on either side of $\gamma$.  The original link carries a Hilbert space $H_\text{in}$, which may be thought of as defined by the `electric' basis $\{|R \rangle \}$ with $R$ ranging over all representations of the gauge group. The entropy of the chosen region is defined in \cite{will14} by taking, for each $R$, the state on the corresponding two half-links to be the maximally-entangled pure state $|\tilde \psi\rangle_{RR}$ on two copies (one for each half-link) of the representation $R$.  The operation mapping the original lattice to the extended lattice of half-links may then be cast in the form \eqref{eq:gencopy} by taking $\{|\alpha \rangle \}= \{|R \rangle \}$ and $| \psi(R)\rangle$ to be supported on some $\text{dim}(R)$ dimensional subspaces of each copy of $H_\text{aux}$, and in which it is unitarily equivalent to $|\tilde \psi\rangle_{RR}$. Here $H_\text{in}$ has dimension equal to the (potentially infinite) number of representations, and the dimension of $H_\text{aux}$ is correspondingly infinite. It is thus natural to apply this version of our edge-mode code when the bulk theory on the links is a lattice gauge theory.

\section{Subsystem Entropy and Edge-Mode Codes}
\label{section:algent}

We now turn to the issue of how our code relates to various discussions of holographic entropy. In particular, following \cite{happy} we show that the entropy of a boundary region $A$ can be written in a form analogous to the explicit terms in the FLM formula \eqref{eq:flm}, but in constrast to the original HaPPY code, for general codes of the form \eqref{eq:gencopy} there is a non-trivial term playing the role of $\frac{\delta \text{Area}}{4G_N}$.  This is to be expected from the algebraic entropy analysis of \cite{harlow}, though we give a direct calculation in section \ref{section:entropy}.  To allow proper comparison, we first review the results of \cite{harlow} before beginning our computation.  We save discussion of the result for section \ref{section:discussion}, where examination of the lattice gauge theory edge-mode codes described below \eqref{eq:gencopy} and comparison with \cite{will14}, \cite{Casini} will suggest that, in a corresponding analysis of linearized gravity, the FLM $\frac{\delta \text{Area}}{4G_N}$ term would be precisely the difference between an extended-lattice bulk entropy analogous to that of \cite{will14} and an algebraic entropy analogous to that of \cite{Casini}.

The general structure of entropy in holographic codes was studied in \cite{harlow}. For an error-correcting code with complementary recovery\footnote{As for \cite{happy}, our code satisfies this requirement for a region $A$ when $\gamma^*_A = \gamma^*_{\bar A}$.} it takes the form

\be
\label{eq:harlow}
S_A= S\left(\rho_A,\mathcal{M}_A\right) + \text{tr}\left(\rho \mathcal{L}_A\right)
\ee
in a code state $\rho$, where $A$ is a subsystem of the bulk Hilbert space. The first term is the entropy of $\rho$ with respect to the von Neumann algebra $\mathcal{M}_A$ associated to $A$, while the second is the expectation value of a linear operator ($\mathcal{L}_A$) lying in the center of $\mathcal{M}_A$.  It was proposed that the $\text{tr}\left(\rho \mathcal{L}_A\right)$ corresponds to the first and third ($\frac{ \text{Area}}{4G_N}$ and  $\frac{\delta \text{Area}}{4G_N}$) terms in \eqref{eq:flm}, while the $S\left(\rho_A,\mathcal{M}_A\right)$ term corresponds to the second ($S(\rho_{W(A)})$) term in \eqref{eq:flm}.

If $\mathcal{M}_A$ has a non-trivial center, any operator in $\mathcal{M}_A$ is block-diagonal, with blocks labeled by eigenvalues $\alpha$ of the center operators:

\be
\label{eq:opdecomp}
\mathcal{O}_A = \bigoplus_\alpha  \mathcal{O}_A^{(\alpha)}.
\ee
This is also true of any state $\rho_A$, which we write as $\rho_A = \bigoplus_\alpha  p_\alpha {\rho}_A^{(\alpha)}$ in terms of normalized states ${\rho}_A^{(\alpha)}$.
The entropy of $\rho_A$ with respect to $\mathcal{M}_A$ can then be defined as $-\text{tr} \left(\rho_A \ln \rho_A \right)$ using the representation \eqref{eq:opdecomp}, so we have

\be
\label{eq:centralent}
S\left(\rho_A,\mathcal{M}_A\right) = -\sum_\alpha p_\alpha \ln p_\alpha + \sum p_\alpha S\left(\rho_A^{(\alpha)}\right),
\ee
with $S\left(\rho_A^{(\alpha)}\right) = -\text{tr} \left(\rho_A^{(\alpha)} \ln \rho_A^{(\alpha)} \right)$.
This is the same as the entanglement entropy we would obtain by tracing the full system state $\rho$ over $\bar{A}$ and then decohering the blocks by deleting all the matrix elements that are not block-diagonal:

\be
S\left(\rho_A,\mathcal{M}_A\right) = S\left(\left[\rho_A\right]_\text{dec}\right)
\ee
where $\left[\rho_A\right]_\text{dec}$ is the reduced density matrix $\rho_A = \text{tr}_{\bar{A}} \rho$ after decohering the blocks.

By contrast, the second contribution to the entanglement in \eqref{eq:harlow} depends on the details of the code.  Since  $\mathcal{L}_A$ lies in the center of $\mathcal{M}_A$, it takes the form

\be
\label{L_A}
\mathcal{L}_A = \bigoplus_\alpha s_\alpha \mathds 1_\alpha.
\ee
$\mathds 1_\alpha$ is the identity matrix on the block $\alpha$ and the $s_\alpha$ are just numbers.

Since the algebras relevant to the original HaPPY code had trivial centers, a primary goal of our work is to manifest this structure in a holographic code.  When the center is trivial there is only one value for the index $\alpha$ and \eqref{eq:centralent} becomes the tautalogy $S(\rho_A)=S(\rho_A)$ and the $\text{tr}(\rho {\mathcal L}_A)$ term becomes independent of the state (though it still depends on $A$).  As reviewed in section \ref{section:entropy} below, in the HaPPY code this constant is given by the logarithm of the bond dimension $\chi$ times the length of the minimal surface anchored to the boundary of $A$.  This term is then interpreted as a model for the leading Ryu-Takayanagi term in the holographic entanglement entropy, so that it is natural to think of $\chi$ as large.  Similarly, the $S(\rho_A,\mathcal{M}_A)$ term is interpreted as modeling the second term in \eqref{eq:flm}.

\subsection{Entropy from Edge Modes}
\label{section:entropy}

We now compute the entropy of an arbitrary boundary region $A$ in our edge-mode code with an arbitrary bulk state.  To the extent that our code satisfies the assumptions of \cite{harlow}, the result must take the form \eqref{eq:harlow}.  The goal of the calculation is thus to determine the explicit form of $\mathcal{L}_A$ for our edge-mode code, as well as to take into account small violations of the complementary recovery assumption of \cite{harlow} (which are also present in the original code \cite{happy}). We will see that $\mathcal{L}_A$ takes the form of a local density on the entangling surface, as appropriate for the term that gives rise to the $\frac{\text{Area}}{4G}$ and $\frac{\delta\text{Area}}{4G}$ pieces of the holographic entanglement.

\begin{figure}[h!]
\centering
\subfloat[]{\includegraphics[width=0.55\textwidth]{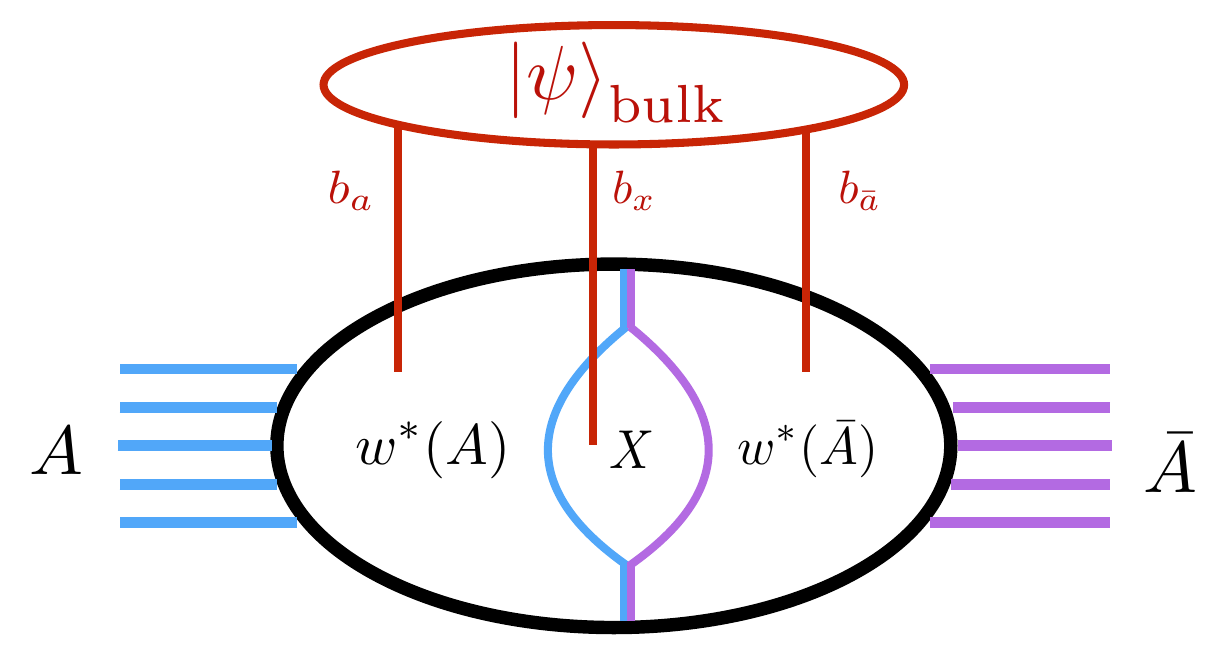}} \qquad
\subfloat[]{\includegraphics[width=0.39\textwidth]{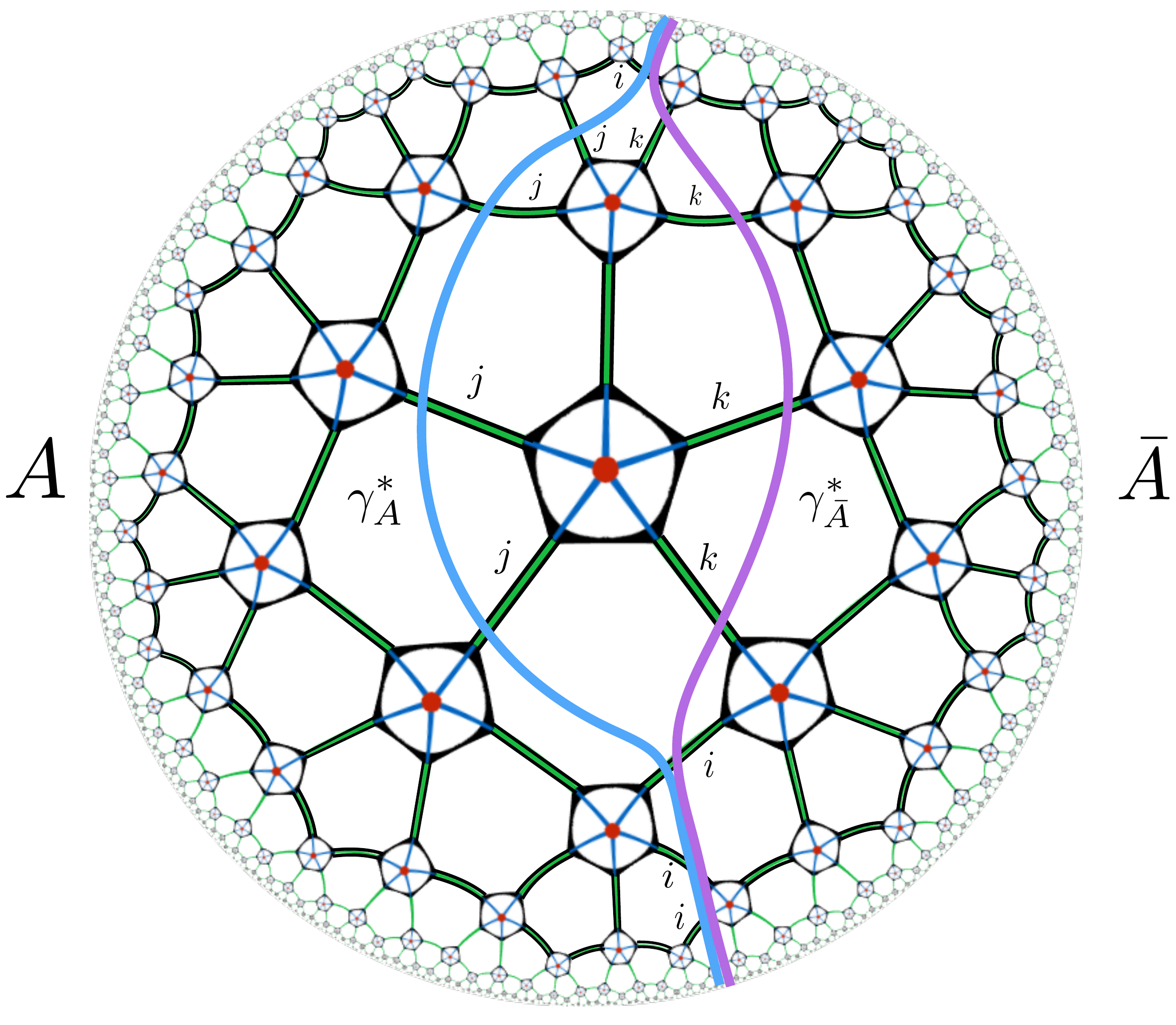}}
\caption{In (a) the network that computes the boundary state, broken into pieces corresponding to the wedges $w^*(A)$, $w^*(\bar A)$, and the residual region $X$. In general all three regions receive inputs from bulk legs. In (b) we label the links cut by $\gamma^*(A)$ and $\gamma^*(\bar A)$ with $j$ and $k$ respectively, and the links cut by both with $i$. 
}
\label{state}
\end{figure}

It is useful to begin by reviewing results for the pentagon code of \cite{happy} on which our edge-mode code is strongly based.  In order to arrive an an explicit FLM-like form, we rewrite the original arguments of \cite{happy} as a tensor network computation.  We begin by thinking of the code as a map from bulk states $|\psi\rangle_{bulk}$ to boundary states $|\psi\rangle_{bndy}$.  In figure \ref{state}, this map is thought of as a tensor network built out of 3 parts: the greedy wedges $w^*(A)$, $w^*(\bar A)$, and a residual bulk region $X = \overline{\left(w^*(A) \cup w^*(\bar A)\right)}$ excluded from both $w^*(A)$ and $w^*(\bar A)$. Here the long overline denotes the complement in the bulk.

\begin{figure}[h!]
\centering
\includegraphics[width=0.9\textwidth]{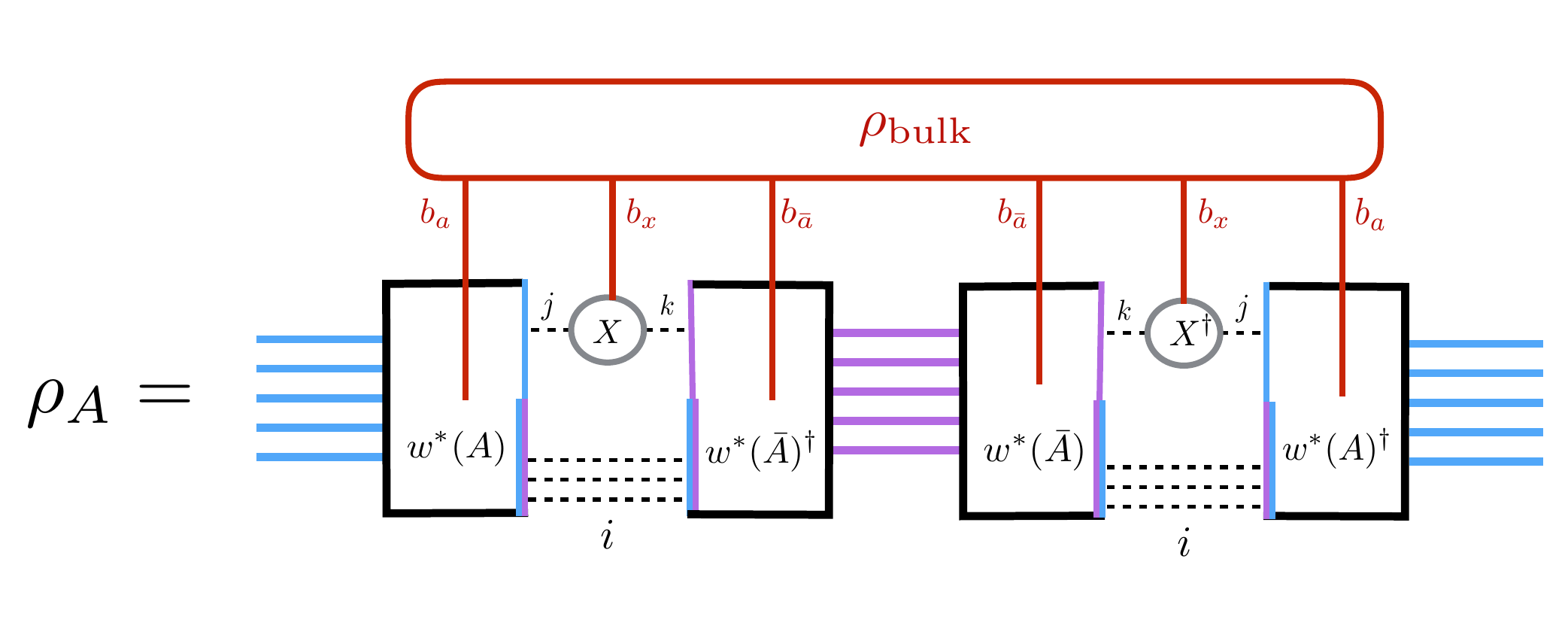}
\caption{The reduced density matrix $\rho_A$ on boundary region $A$ described as a circuit and broken into pieces corresponding to $w^*(A)$, $w^*(\bar A)$, and $X$.  The $j$ links describe network edges cut by $\gamma_{A}^*$,  the $k$ links describe network edges cut by $\gamma_{\bar A}^*$, and the $i$ links describe those cut by both.
}
\label{density}
\end{figure}

\begin{figure}[h!]
\centering
\includegraphics[width=0.6\textwidth]{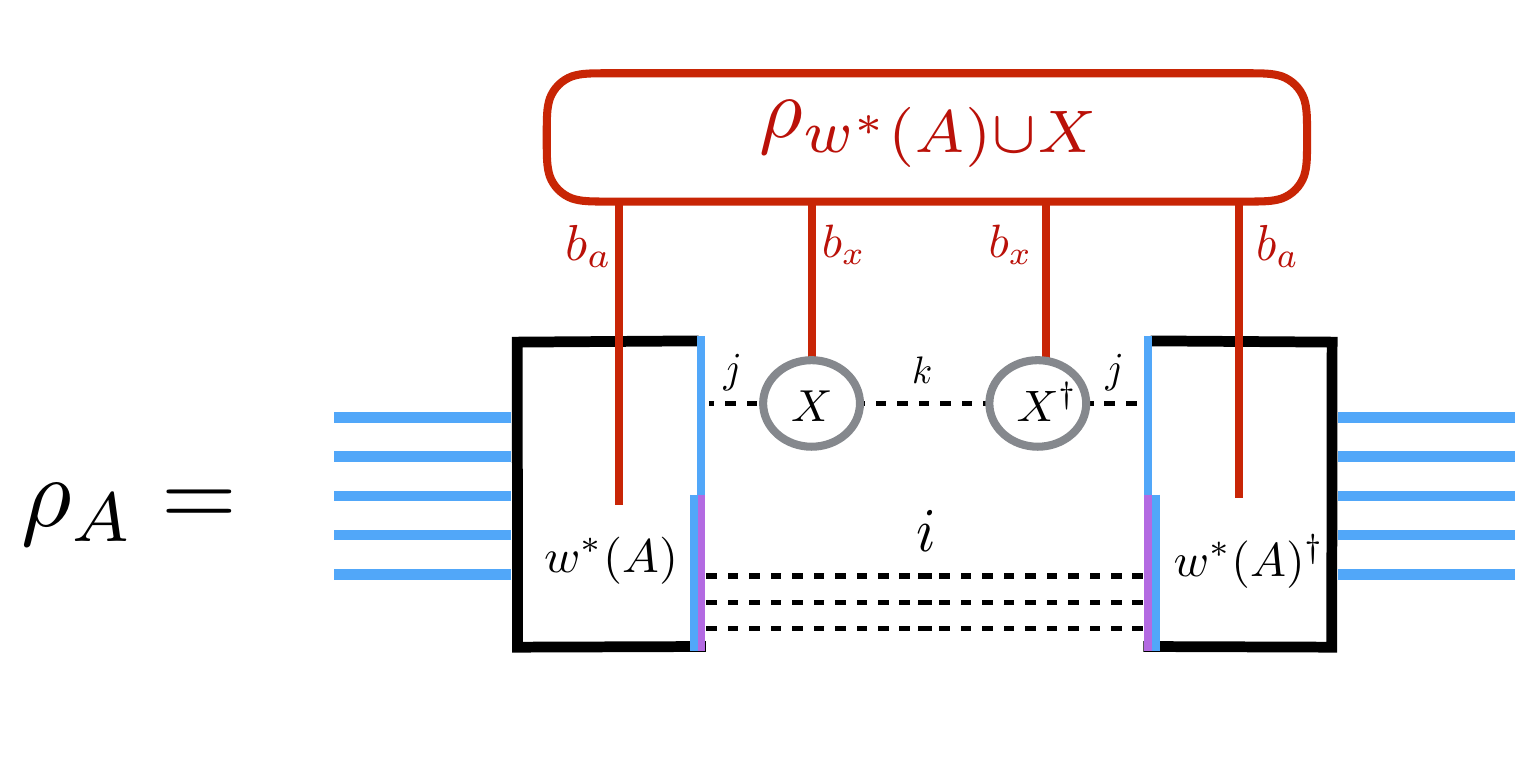}
\caption{A simpler network for $\rho_A$ obtained from figure \ref{density} by using the fact that $w^*(\bar A)$ defines an isometry from the indices $k$, $i$, and $b_{\bar a}$ to the boundary $\bar A$.
}
\label{simp1}
\end{figure}

We may of course also use this network to map bulk density matrices $\rho_{bulk}$ to boundary density matrices $\rho_{bndy}$.  Although $\rho_{bndy}$ is defined on the entire boundary, tracing over $\bar A$ gives a reduced density matrix $\rho_A$ on boundary region $A$. The associated tensor network is shown in figure \ref{density}, displaying the different roles performed by network links cut by $\gamma_A^*$, those cut by $\gamma_{\bar A}^*$, and those cut by both.
We then recall that $w^*(\bar A)$  defines an isometry (which we will also denote as $w^*(\bar A)$) from the associated bulk indices and the network edges cut by $\gamma^*_{\bar A}$ to $\bar A$, i.e. it satisfies
\begin{equation}
[w^*(\bar A)]^\dagger w^*(\bar A) = \mathds{1},
\end{equation}
with $\mathds{1}$ being the identity on the space of inputs.  As a result, the tensor network may be simplified to that shown in figure \ref{simp1}.

\begin{figure}[h!]
\centering
\includegraphics[width=0.48\textwidth]{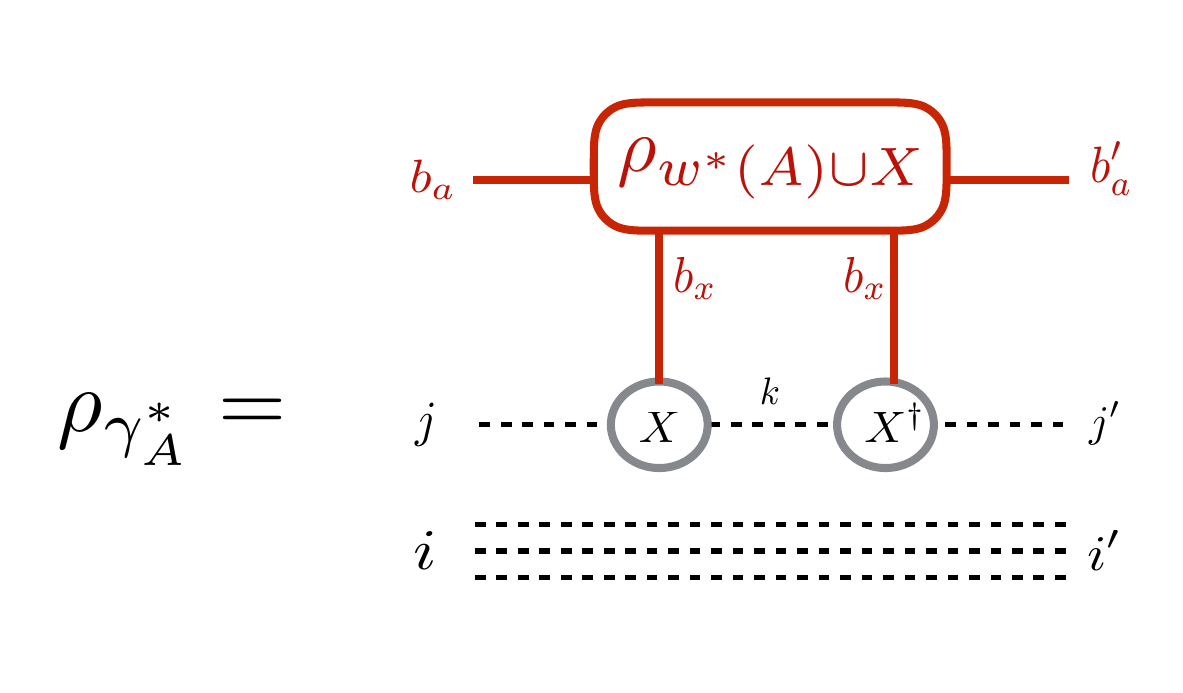}
\caption{The state $\rho_{\gamma_A^*}$ used to compute $S_A$. 
}
\label{simp2}
\end{figure}

A key part of this network representation of $\rho_A$ is given by the state $\rho_{\gamma_A^*}$, defined on the Hilbert space $H_{w^*(A)} \otimes  H_{\gamma^*_{A}}$ as show in figure \ref{simp2}. Here $ H_{w^*(A)}$ is the space associated with bulk inputs to $w^*(A)$, and $H_{\gamma^*_{A}}$ is the space defined by network edges cut by $\gamma^*_A$.
Figure \ref{simp1} then implies that we may write
\ban{
\rho_A =  w^*(A) \rho_{\gamma_A^*}  w^*(A)^\dag\, 
}
in terms of the isometry $w^*(A)$. Since isometries preserve von Neumann entropy, the entropy of the boundary region $A$ may be written
\begin{equation}
S_A = - \text{tr} \left( \rho_A \ln \rho_A\right) = - \text{tr} \left( \rho_{\gamma^*_A} \ln \rho_{\gamma^*_A}\right).
\end{equation}
To the extent that we can ignore the excluded region $X$, we have that $\rho_{w^*(A) \cup X} \approx \rho_{{w^*( A)}}$ and the density matrix $\rho_{\gamma^*_A}$ is
\ban{
\rho_{\gamma^*_A} \sim \rho_{{w^*(\bar A)}} \otimes \mathds{1}_i \otimes \mathds{1}_j
}
up to normalization. The von Neumann entropy $S_A$ is then precisely
\begin{equation}
\label{eq:ignoreX}
S_A = S(\rho_{w^*(A)}) + |\gamma^*_{A}|\ln\chi \, ,
\end{equation}
in terms of the bond dimension $\chi$ of each network edge and the number $|\gamma^*_{A}|$ of edges cut by $\gamma^*_{A}$ (i.e. the total number of $i$ and $j$ indices).

More generally, the region $X$ introduces further corrections to
 \eqref{eq:ignoreX}.   While such corrections are difficult to compute explicitly, subadditivity and the Araki-Lieb inequality ($|S_{B} - S_C| \le S_{BC} \le S_B + S_C$) can be used to bound departures from this estimate in terms of $\chi$ and the number of $j$ edges as defined in figure \ref{density} (those cut by $\gamma^*_A$ but not by $\gamma^*_{\bar A}$). In special cases $X$ can be quite large, but this is not generally the case \cite{happy}.  Indeed, $X$ can often be made to vanish by moving a small number of boundary points from $A$ to $\bar A$ and/or from $\bar A$ to $A$.  It is thus natural to think of think of $X$ as an artifact of the discrete toy model used here.  As a result, although there is no limit of our model in which the region $X$ can be systematically neglected in all cases, as in \cite{happy} we choose to ignore the region $X$ when considering implications for gauge/gravity duality. We thus consider only cases with trivial $X = \emptyset$ below.

With this technology in hand, we now turn to our edge mode code. Our code attaches bulk states $\ket{\psi}_\text{bulk, EMC}$ to a product of HaPPY codes via an isometry $\mathscr G$ built from the copying tensors $G$. The end result is thus just what would be obtained by feeding the state
\be
\ket{\psi}_\text{bulk, HaPPY}=\mathscr{G}\ket{\psi}_\text{bulk, EMC} 
\ee
into a product of HaPPY codes. Thus all that remains is to replace $\rho_{w^*(A) \cup X}$ in figure \ref{simp2} with the corresponding density matrix obtained from the state with $G$s inserted.

Now, since $G$ is an isometry from its single input to its pair of outputs, adding $G$-tensors with both outputs in $w^*(A)$ will not change the entropy of figure \ref{simp2} and may be ignored.  Furthermore, the trace over $w^*(\bar A)$ in passing from figure \ref{state} to figure \ref{simp1} removes all $G$'s with both outputs in $w^*(\bar A)$. When $X=\emptyset$, $G$ has no outputs in $X$.  
 
\begin{figure}[h!]
\centering
\includegraphics[width=0.5\textwidth]{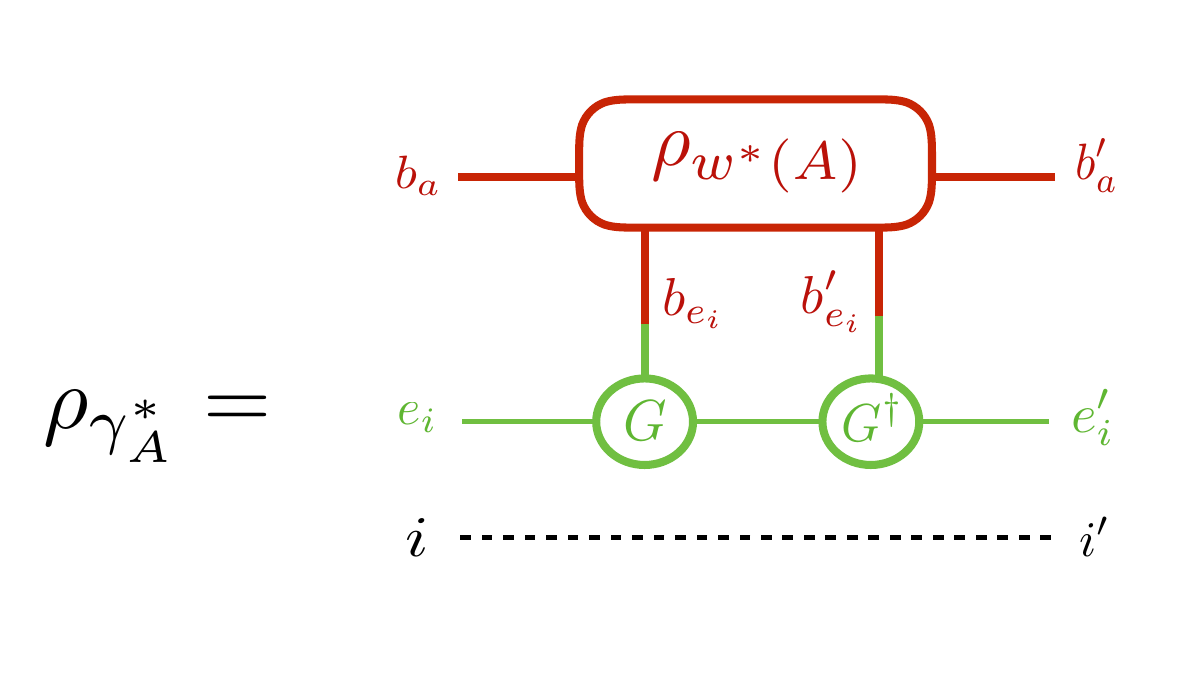}
\caption{A state $\rho_{\gamma_A^*}$ that may be used to compute $S_A$ in our edge-mode code when $X=\emptyset$. Every leg $i$ cut by $\gamma_A^*$ has an associated $b_{e_i}$ and $e_i$, although only one such leg is drawn for simplicity. 
}
\label{withedge}
\end{figure}

Thus the only remaining $G$s to consider are those with one output in each of $w^*(A)$, $w^*(\bar A)$.  They will act on bulk indices which we may organize in pairs containing one unprimed index (a ket index of $\rho_{w^*(A)}$) and the corresponding primed index (a bra index of $\rho_{w^*(A)}$).  As shown in figure \ref{withedge}, the $w^*(\bar A)$ output legs of the pair of $G$-tensors acting on these bulk indices are contracted by the above-mentioned trace over $w^*(\bar A)$.

\begin{figure}[h!]
\centering
\includegraphics[width=0.75\textwidth]{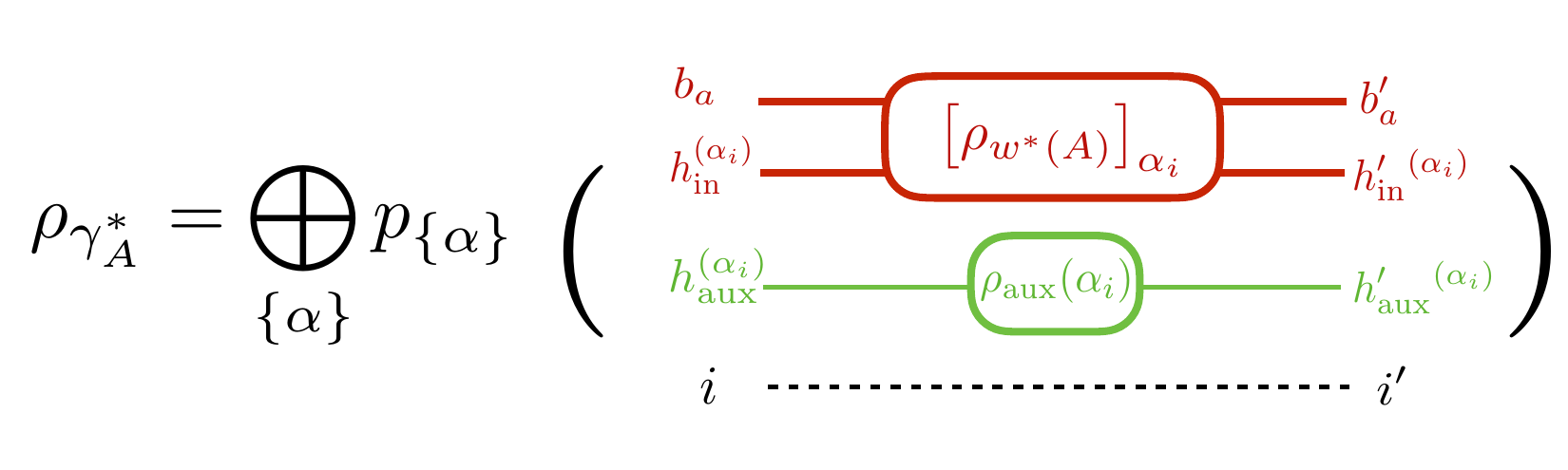}
\caption{The state $\rho_{\gamma_A^*}$ in terms of the states $\left[\rho_{\gamma_A^*}\right]_{\{\alpha\}}$ obtained by decohering the tensors $G$ in figure \ref{withedge} with respect to the $\alpha$ basis. The index $h_\text{in}^{(\alpha_i)}$ refers to the subspace of $(H_\text{in})_L$ at link $i$ associated with the eigenvalue $\alpha$, and similarly for $h_\text{aux}^{(\alpha_i)}$. Again we have drawn only one link $i$ for simplicity. 
}
\label{alphastate}
\end{figure}

The effect of these final $G$s on the entropy is easy to understand by thinking about the action of a single $G$ on any pure-state input $|\psi \rangle = \sum_\alpha c_\alpha |\alpha \rangle$.  Due to the copying property of $G$ in \eqref{eq:gencopy}, the action of $G$ on $|\psi\rangle$ may be thought of as a von Neumann measurement; i.e., as a unitary transformation that entangles the original system (here $(H_\text{in})_L$, which we take to lie in $w^*(A)$) with a `measuring apparatus' $(H_\text{in} \otimes H_\text{aux})_R$ in $w^*(\bar A)$.  At the same time, it also creates further entanglement with $(H_\text{aux})_L$.  Tracing over the $w^*(\bar A)$-output then decoheres the state into the copying basis $\{\alpha \}$ so that $\rho_{\gamma_A^*}$ is block-diagonal in this basis, and the entanglement with $(H_\text{aux})_L$ means that the $|\alpha \rangle \langle \alpha |$-blocks appear tensored with the state
\begin{equation}
\rho_\text{aux} (\alpha)= \text{tr}_{(H_\text{aux})_R} \left( |\psi(\alpha) \rangle \langle \psi(\alpha) | \right).
\end{equation}
on $(H_\text{aux})_L$. Thus we have
\begin{equation}
\rho_{\gamma_A^*} = \bigoplus_{\{\alpha\}} p_{\{\alpha\}} \left(\left[\rho_{\gamma_A^*}\right]_{\{\alpha\}} \otimes \rho_\text{aux}({\{\alpha\}}) \right),
\label{decstate}
\end{equation}
with $p_{\{\alpha\}} = \prod_i |c_{\alpha_i}|^2$, 
 $\rho_\text{aux} ({\{\alpha\}}) = \otimes_i \ \rho_\text{aux}(\alpha_i)$,
 and $i$ again ranging over all links cut by both $\gamma^*_A$ and $\gamma^*_{\bar A}$.  The relevant $\left[\rho_{\gamma_A^*}\right]_{\{\alpha\}}$ may be described by introducing the $\alpha$-decohered bulk state
\begin{equation}
\label{decwedge}
\left[ \rho_{w^*(A)}  \right]_\text{dec} = \bigoplus_{\{\alpha\}} p_{\{\alpha\}} \left[\rho_{w^*(A)}\right]_{\{\alpha\}}.
\end{equation}
The $\left[\rho_{\gamma_A^*}\right]_{\{\alpha\}}$ are then
given by figure \ref{simp2} (here with $X=\emptyset$ and thus no $j,j'$ indices) in terms of the $\left[\rho_{w^*(A)}\right]_{\{\alpha\}}$ defined by \eqref{decwedge}. Putting everything together, we draw the network representation of \eqref{decstate} in figure \ref{alphastate}. 

This gives
\begin{eqnarray}
\label{eq:Swithedge}
S_A &=&  |\gamma^*_{A}|\ln\chi  + \sum_{\{\alpha\}}   p_{\{\alpha\}} \left( - \ln p_{\{\alpha\}} +   S(\left[\rho_{w^*(A)}\right]_{\{\alpha\}}) + S\left(\rho_\text{aux}({\{\alpha\}})\right) \right) \, \cr
&=&  S(\rho_A, \mathcal{M}_A) + |\gamma^*_{A}|\ln\chi  + \sum_{\{\alpha\}}   p_{\{\alpha\}} S\left(\rho_\text{aux} ({\{\alpha\}})\right) \,
,
\end{eqnarray}
where $\chi$ now denotes the total bond dimension associated with the $T$ tensors (i.e. $6$ times that of the HaPPY code for the code described in section \ref{section:construction}). Here we have used \eqref{eq:centralent} to identify the ``bulk entanglement'' term $S(\rho_A,M)$.  Note that \eqref{eq:Swithedge} takes Harlow's form \eqref{eq:harlow} if we define $\mathcal{L}_A$ by \eqref{L_A} and make the further identification
\begin{equation}
s_{\{\alpha\}} = |\gamma^*_{A}| \ln\chi + S\left(\rho_\text{aux} ({\{\alpha\}})\right) =\sum_i \left[ \ln \chi + S\left(\rho_\text{aux} ({\alpha_i})\right)\right].
\label{eq:scoeff}
\end{equation}
This manifestly takes the form of a local density on the entangling surface. The contribution of the first term in \eqref{eq:scoeff} to $S_A$ is independent of the state, analogous to the leading Ryu-Takayanagi piece in the entropy. The contribution of the second piece to $S_A$ linearly second depends on the bulk state, like the $\frac{\delta\text{Area}}{4G}$ correction of \cite{Faulkner:2013ana}.

As a concrete example, consider the copying tensor defined in \eqref{eq:gravitoncode}. It is easy to show that $s_0 = 0$ and $s_1 = \ln 2$.

\section{Discussion}
\label{section:discussion}

We have constructed edge-mode holographic codes by composing (copies of) the HaPPY pentagon code \cite{happy} with certain `copying tensors' $G$ \eqref{eq:gencopy}.  The results provide toy models for holography that implement the structure described in \cite{harlow}.  In particular, subregions $A$ of the boundary allow the reconstruction of bulk algebras $M$ having a non-trivial center associated with the interior boundary of the (greedy) entanglement wedge $w^*(A)$, i.e. with the holographic code analogue of the Ryu-Takayanagi minimal surface.  As a result, subject to the same caveats as for the original HaPPY code \cite{happy}, our model gives rise to an FLM-like relation \eqref{eq:harlow}.   We expect that a similar edge-mode extension can be applied to other holography-inspired codes including \cite{Qi:2013caa,Yang:2015uoa,haydenRTN}.

In particular, the linear operator $\mathcal{L}_A$ of \cite{harlow} receives a contribution that depends on the choice of copying tensor $G$.  For general $G$ this term depends non-trivially on the bulk state.  This behavior is in contrast to that of contributions from the $|\gamma^*_{A}|\ln  \chi$ term in $\chi_{\{\alpha\}}$.  Because the $ |\gamma^*_{A}|\ln  \chi$  term does not depend on $\alpha$, it contributes $\sum_{\{\alpha\}} p_{\{\alpha\}} |\gamma^*_{A}| \ln  \chi= |\gamma^*_{A}| \ln  \chi$ to the entropy for any bulk state.

Due to this distinction, and
following the spirit of \cite{harlow}, it is natural to think of the $|\gamma^*_{A}| \ln  \chi$ term in \eqref{eq:Swithedge} as modeling the Ryu-Takayanagi term in \eqref{eq:flm}, the
$S(\rho_A, \mathcal{M}_A)$ term as corresponding to $S_{bulk}(\rho_{w(R)})$, and the remaining term $\sum_{\{\alpha\}}   p_{\{\alpha\}} S\left(\rho_\text{aux}({\{\alpha\}})\right)$ from $\text{tr} \left(\rho \mathcal{L}_A \right)$ as modeling FLM's $\frac{\delta \text{Area}}{4G_N}$.    We expect that this is roughly correct, though it remains to be verified in detail due to a subtle difference between \cite{harlow} and the approach of FLM.   The point here is that to identify a part of $\text{tr} \left(\rho \mathcal{L}_A \right)$ with $\frac{\delta \text{Area}}{4G_N}$ we must take the bulk state to include propagating metric fluctuations.  But the analysis \cite{Faulkner:2013ana} of FLM treated the bulk as a `normal' quantum field theory which could be defined on any metric background, and in particular on backgrounds with conical defects.  This is not the case for metric fluctuations which propagate consistently only when the background is on-shell, and so a conclusive result awaits a more complete re-analysis of FLM including a careful treatment of bulk gravitons.\footnote{A further source of confusion, though not real difficulty, is the fact that \cite{harlow} allows an arbitrary bulk state to be considered independent of any choice of background, while the semi-classical approach of  \cite{Faulkner:2013ana} naturally correlates the bulk state with the background.  In particular, this approach generally selects bulk state in which deviations from the background metric have vanishing expectation value, so that there is no explicit first-order contribution to $\frac{\delta \text{Area}}{4G_N}$ from linearized gravitons and, instead, this term receives contributions only from back-reaction and quadratic terms at the next order.  But the freedom to expand around a different background makes clear that, in principle, this $\frac{\delta \text{Area}}{4G_N}$ would indeed receive a linear contribution from linearized metric fluctuations.}

However, perturbative gravity is a gauge theory having much in common with Yang-Mills theory.  It is thus interesting to consider in detail the form of \eqref{eq:Swithedge} when the bulk degrees of freedom are taken to describe Yang-Mills. As described at the end of section \ref{section:reconstruction}, it is natural to do so by introducing a lattice gauge theory on the links of the network links and attaching the bulk state to $5$ pentagon codes using copying tensors that transform this lattice into an extended lattice as in \cite{will_lattice,will14}.
Only $5$ pentagon codes are required as, for the moment, we suppose that the gauge theory defines all bulk degrees of freedom.

The resulting system can now be viewed in two different ways.  The viewpoint used thus far is that the bulk system consists of lattice gauge theory on the network links and that we act on this system with our edge-mode code to obtain the associated boundary state.  This leads to \eqref{eq:Swithedge} and the above identification of terms.  However, the result can equally-well be viewed as a bulk system defined by an {\it extended} lattice, in which each link has been replaced by a pair of half-links, acted on by a code that is precisely the tensor product of $5$ copies of the HaPPY pentagon code.    We may then use the result \eqref{eq:ignoreX}, taking the first term on the right to be the entropy $S(\rho^{ext}_{w^*(A)})$ defined by the extended bulk lattice.  The entropy of such extended lattice states was discussed in \cite{will14}, and letting $\alpha_i$ range over representations $R_i$ of the gauge group as at the end of section \ref{section:reconstruction}, takes the form
\begin{eqnarray}
\label{extlat}
S(\rho^{ext}_{w^*(A)}) &=& 
\sum_{\{R\}} p_{\{R\}} \left( -  \ln p_{\{R\}} + S(\left[\rho_{w^*(A)}\right]_{\{R\}}) + \sum_i \ln (\text{dim}\ R_i)  \right) \cr
&=& S(\rho_A, \mathcal{M}_A)  + 
\sum_{\{\alpha\}} p_{\{\alpha\}} \sum_i S\left(\rho_\text{aux}(\alpha_i) \right), 
\end{eqnarray}
as expected for agreement with our previous computation of $S_A$. 

The final result \eqref{extlat} becomes particularly interesting if we assume that corresponding results for perturbative gravitons are given by a naive extrapolation. Given the identification in that context of the FLM $\frac{\delta \text{Area}}{4G_N}$ term with the second term on the right hand side of \eqref{extlat} as described above, such extrapolation suggests that -- at least to some order in the bulk Newton constant -- the FLM relation may be rewritten as simply
\begin{equation}
\label{extFLM}
S_A = \frac{\text{Area}}{4G} + S(\rho^{ext}_{w(A)}),
\end{equation}
with the first term computed in some classical background and the second defined by an appropriate extended lattice construction for the perturbative metric fluctuations.  The form \eqref{extFLM} is particularly natural given the reliance of FLM on the replica trick and the agreement between the replica trick and extended lattice constructions noted in \cite{Donnelly:2014fua,Donnelly:2015hxa}.  It would thus be very interesting to explore such a construction directly in linearized gravity, either on a lattice or in the continuum with an appropriate corresponding extension of the Hilbert space, and to establish any relation to the fully non-linear extended classical phase space for gravity described recently in \cite{Donnelly:2016auv}.


\section*{Acknowledgements}
It is a pleasure to thank Horacio Casini, Zicao Fu, Brianna Grado-White, Dan Harlow, Veronika Hubeny, Aitor Lewkowycz, and Mukund Rangamani for useful conversations. DM was supported in part by the Simons Foundation and by funds from the University of California. BM was supported by NSF Grant PHY13-16748. This research was supported in part by Perimeter Institute for Theoretical Physics. Research at Perimeter Institute is supported by the Government of Canada through the Department of Innovation, Science and Economic Development and by the Province of Ontario through the Ministry of Research and Innovation. 


\bibliographystyle{jhep}
\phantomsection
\renewcommand*{\bibname}{References}

\bibliography{references}

\providecommand{\href}[2]{#2}\begingroup\raggedright\begin{thebibliography}{10}

\bibitem{almheiri}
A.~Almheiri, X.~Dong, and D.~Harlow, {\it {Bulk Locality and Quantum Error
  Correction in AdS/CFT}},  {\em JHEP} {\bf 04} (2015) 163,
  [\href{http://arxiv.org/abs/1411.7041}{{\tt arXiv:1411.7041}}].

\bibitem{happy}
F.~Pastawski, B.~Yoshida, D.~Harlow, and J.~Preskill, {\it {Holographic quantum
  error-correcting codes: Toy models for the bulk/boundary correspondence}},
  {\em JHEP} {\bf 06} (2015) 149, [\href{http://arxiv.org/abs/1503.06237}{{\tt
  arXiv:1503.06237}}].

\bibitem{haydenRTN}
P.~Hayden, S.~Nezami, X.-L. Qi, N.~Thomas, M.~Walter, and Z.~Yang, {\it
  {Holographic duality from random tensor networks}},
  \href{http://arxiv.org/abs/1601.01694}{{\tt arXiv:1601.01694}}.

\bibitem{harlow}
D.~Harlow, {\it {The Ryu-Takayanagi Formula from Quantum Error Correction}},
  \href{http://arxiv.org/abs/1607.03901}{{\tt arXiv:1607.03901}}.

\bibitem{rt1}
S.~Ryu and T.~Takayanagi, {\it {Holographic derivation of entanglement entropy
  from AdS/CFT}},  {\em Phys. Rev. Lett.} {\bf 96} (2006) 181602,
  [\href{http://arxiv.org/abs/hep-th/0603001}{{\tt arXiv:hep-th/0603001}}].

\bibitem{rt2}
S.~Ryu and T.~Takayanagi, {\it {Aspects of Holographic Entanglement Entropy}},
  {\em JHEP} {\bf 08} (2006) 045,
  [\href{http://arxiv.org/abs/hep-th/0605073}{{\tt arXiv:hep-th/0605073}}].

\bibitem{Faulkner:2013ana}
T.~Faulkner, A.~Lewkowycz, and J.~Maldacena, {\it {Quantum corrections to
  holographic entanglement entropy}},  {\em JHEP} {\bf 11} (2013) 074,
  [\href{http://arxiv.org/abs/1307.2892}{{\tt arXiv:1307.2892}}].

\bibitem{Buividovich}
P.~V. Buividovich and M.~I. Polikarpov, {\it {Entanglement entropy in gauge
  theories and the holographic principle for electric strings}},  {\em Phys.
  Lett.} {\bf B670} (2008) 141--145,
  [\href{http://arxiv.org/abs/0806.3376}{{\tt arXiv:0806.3376}}].

\bibitem{will_lattice}
W.~Donnelly, {\it {Decomposition of entanglement entropy in lattice gauge
  theory}},  {\em Phys. Rev.} {\bf D85} (2012) 085004,
  [\href{http://arxiv.org/abs/1109.0036}{{\tt arXiv:1109.0036}}].

\bibitem{Casini}
H.~Casini, M.~Huerta, and J.~A. Rosabal, {\it {Remarks on entanglement entropy
  for gauge fields}},  {\em Phys. Rev.} {\bf D89} (2014), no.~8 085012,
  [\href{http://arxiv.org/abs/1312.1183}{{\tt arXiv:1312.1183}}].

\bibitem{will14}
W.~Donnelly, {\it {Entanglement entropy and nonabelian gauge symmetry}},  {\em
  Class. Quant. Grav.} {\bf 31} (2014), no.~21 214003,
  [\href{http://arxiv.org/abs/1406.7304}{{\tt arXiv:1406.7304}}].

\bibitem{Qi:2013caa}
X.-L. Qi, {\it {Exact holographic mapping and emergent space-time geometry}},
  \href{http://arxiv.org/abs/1309.6282}{{\tt arXiv:1309.6282}}.

\bibitem{Yang:2015uoa}
Z.~Yang, P.~Hayden, and X.-L. Qi, {\it {Bidirectional holographic codes and
  sub-AdS locality}},  {\em JHEP} {\bf 01} (2016) 175,
  [\href{http://arxiv.org/abs/1510.03784}{{\tt arXiv:1510.03784}}].

\bibitem{Donnelly:2014fua}
W.~Donnelly and A.~C. Wall, {\it {Entanglement entropy of electromagnetic edge
  modes}},  {\em Phys. Rev. Lett.} {\bf 114} (2015), no.~11 111603,
  [\href{http://arxiv.org/abs/1412.1895}{{\tt arXiv:1412.1895}}].

\bibitem{Donnelly:2015hxa}
W.~Donnelly and A.~C. Wall, {\it {Geometric entropy and edge modes of the
  electromagnetic field}},  \href{http://arxiv.org/abs/1506.05792}{{\tt
  arXiv:1506.05792}}.

\bibitem{Donnelly:2016auv}
W.~Donnelly and L.~Freidel, {\it {Local subsystems in gauge theory and
  gravity}},  {\em JHEP} {\bf 09} (2016) 102,
  [\href{http://arxiv.org/abs/1601.04744}{{\tt arXiv:1601.04744}}].

\end{thebibliography}\endgroup

\end{document}